\documentclass[aps,prl,amsfonts,amssymb,showpacs,graphicx,epsfig,twocolumn]{revtex4}
\usepackage{amssymb}
\usepackage[dvips]{graphicx}
\usepackage{epsfig}
\draft
\begin{document}
\title{Giant Wave-Drag Enhancement of Friction in Sliding Carbon Nanotubes}
\author{Paul Tangney$^{1,2}$, Marvin L. Cohen$^{2,3}$, and Steven G. Louie$^{1,2,3}$}
\date{\today}
\date{June 25, 2006}


\affiliation{
$^{1}$The Molecular Foundry, Lawrence Berkeley National Laboratory, CA 94720\\
$^{2}$Materials Sciences Division, Lawrence Berkeley National Laboratory, CA 94720.\\
$^{3}$Department of Physics, University of California, Berkeley, CA 94720}

\begin{abstract}
Molecular dynamics simulations of coaxial carbon nanotubes
in relative sliding motion reveal a striking enhancement of 
friction when phonons whose group velocity is close 
to the sliding velocity of the nanotubes are resonantly excited. 
The effect is analogous to the dramatic increase in air drag 
experienced by aircraft flying close to the speed of sound, but
differs in that it can occur in multiple velocity ranges with 
varying magnitude, depending on the atomic level structures of the
nanotubes. The phenomenon is a general one that may occur in other
nanoscale mechanical systems.
\end{abstract}
\maketitle

The miniaturization of mechanical devices has evolved to the point 
that microelectromechanical systems (MEMS) are now in widespread use
in a growing number of commercial products.
Miniaturization has many advantages including 
increased speed and sensitivity, reduced power consumption, and 
improvements in the positional and spatial precision with which 
complex functions can be performed.
At all length scales the scale-relative properties of mechanical systems vary 
with their size due to increasing edge to surface and surface to volume 
ratios. However, as the nanometer 
length scale is approached, an entirely new set of phenomena are 
becoming relevant that provide great challenges to the 
further miniaturization of devices. The precise atomic and electronic 
structure of devices, the chemical reactivity of device components, 
the quantum mechanical behaviour of atoms, and the presence of large 
thermal fluctuations are all issues of increasing importance at this scale.
Engineering useful nanoscale mechanical devices will require an understanding of 
these effects and their impacts on device efficiency and reliability.

Multiwalled carbon nanotubes (MWCNTs) can be fabricated with 
near structural perfection and their resistance to deformation
and the smoothness with which coaxial tubes can move relative to 
one another suggest that they could be ideal materials from which 
to build nanoscale machinery.
While {\em useful} nano-mechanical systems have
not yet been constructed, some prototype mechanical elements with sub-micron 
dimensions have been made using MWCNTs\cite{cumingsetc} and 
it seems unlikely that the ultimate lower size limit of functional mechanical 
devices has, on the micron scale, already been reached.
Theory and simulation can help to investigate the limitations of nanoscale 
machinery and to improve their design by providing a more detailed understanding of 
energy loss mechanisms than can be inferred from experiment.
This Letter is concerned with one of the most important 
gross features of nanoscale phononic friction: Its dependence on velocity.

\begin{figure}
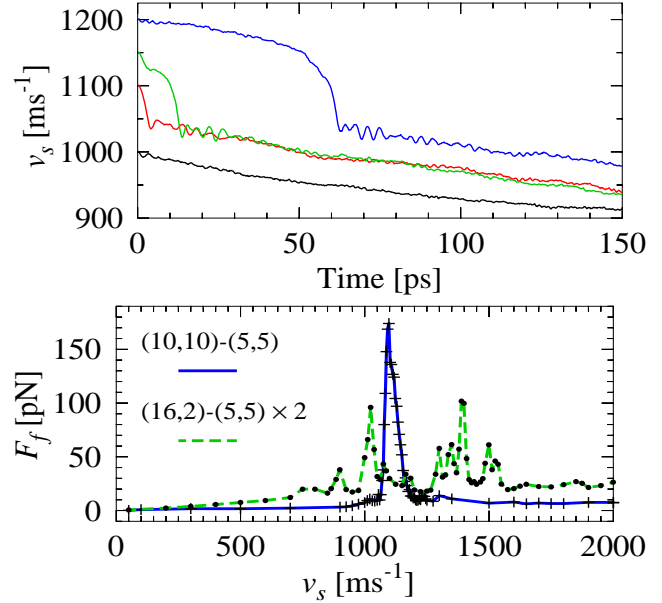
 
\epsfig{figure=fig1a.eps,height=4.5cm,width=8.5cm,height=4cm}
\epsfig{figure=fig1b.eps,height=4.5cm,width=8.5cm,height=4cm}
\caption{(a) Sliding velocity, $v_s$, as a function of time for four different
molecular dynamics (MD) simulations of a (5,5) CNT sliding within a (10,10) CNT.
(b) Average friction force, $F_f$, as a function of the average
sliding velocity in near-constant-velocity MD simulations. For the (16,2)-(5,5) system, $F_f$ has
been scaled by a factor of two for visibility.}
\label{fig1}
\end{figure}
\begin{figure}
\epsfig{figure=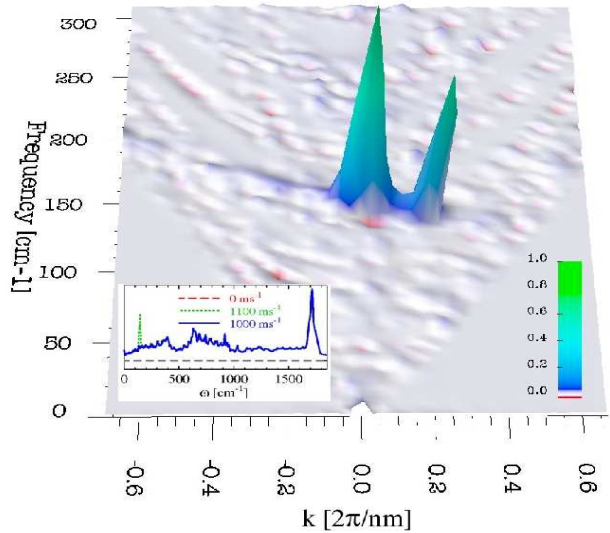,width=8.5cm,height=7.5cm}
\caption{
The {\em difference} in $(\omega,k)$ space
between the power spectra of the (10,10) CNT calculated 
at sliding velocities of $1100 \;\text{ms}^{-1}$ and 
$1000 \;\text{ms}^{-1}$. Inset: Frequency-dependent
power spectrum at different sliding velocities.}
\label{fig2}
\end{figure}
\begin{figure}
\epsfig{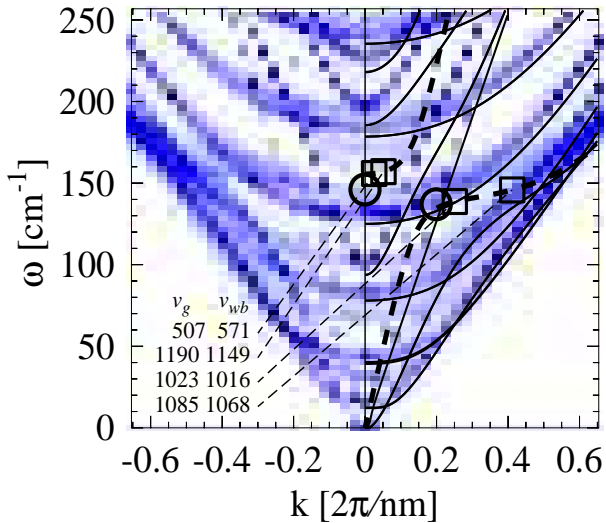}
\caption{The lines are the phonon bands of an isolated $(10,10)$ nanotubes
at $0$ K. The RB and LA modes are indicated by dashed lines.
The underlying color plot is the $k$-resolved phonon power spectrum of the $(10,10)$
CNT when the $(5,5)$ shuttle is moving inside it at $1000 \;\text{ms}^{-1}$.
Circles indicate the positions of the peaks seen in Fig.\ref{fig2}
while the boxes are points that satsify the proposed conditions
for enhanced friction.}
\label{fig3}
\end{figure}
\begin{figure}
\epsfig{figure=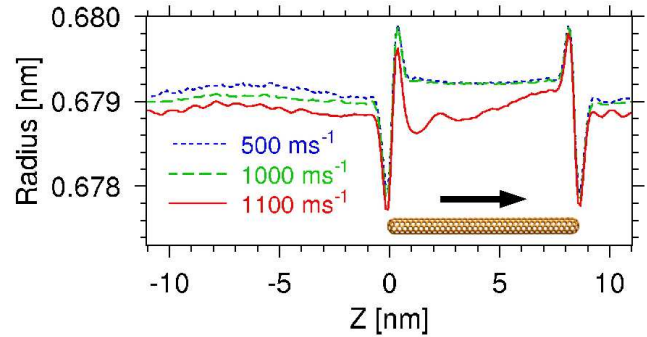,width=8.5cm}
\caption{Average over time and azimuthal angle of the local radius, 
$r_{\text{ave}}$ of the $(10,10)$ CNT as a function of 
displacement from one end of the $(5,5)$ CNT at
three different sliding velocities}
\label{fig4}
\end{figure}

Small volume (mass) to surface (interaction area) ratios\cite{drexler,us}
mean that components of nanoscale devices can move with enormous
accelerations and velocities. Simulations of an 
oscillator composed of double-walled carbon nanotubes(CNT)\cite{zheng} 
showed that even the weak van der Waals interactions between
layers of a MWCNT are sufficient to accelerate nanotubes 
to velocities of thousands of meters per second\cite{us}.

In this work, some results of molecular dynamics (MD) simulations of
CNT ``shuttles'', of varying length, sliding within a larger radius 
CNT under periodic boundary conditions\cite{technical_details}
are reported.
The focus is on a striking phenomenon, involving an enhancement of friction
of several orders of magnitude, that occurs in certain velocity
ranges and that may have general importance for the design of
nanomechanical systems\cite{us_aps}.
Figure \ref{fig1}(a) is a plot of sliding velocity, $v_s$, as a function 
of time in MD simulations of a $8$ nm $(5,5)$ CNT shuttle
sliding within a $(10,10)$ CNT in a periodically repeated unit cell 
of length $\sim 22.1$ nm\cite{length_check1}.
When $v_s$ is in the approximate range $1050 \;\text{ms}^{-1}$ 
to $1150 \text{ms}^{-1}$, the rate at which the shuttle slows down is 
dramatically larger than when $v_s$ is outside this range.

To investigate this effect, MD simulations have been performed at 
close to constant velocity as follows:
With the CNTs at rest, the system was prepared at a temperature
of $300$ K and then equilibrated in a microcanonical ensemble
for tens of picoseconds.
The shuttle was then given an initial sliding 
velocity, $v_s^0$, along the common axis of the nanotubes and MD was performed
in a microcanonical ensemble, but with $v_s$ rescaled to $v_s^0$ every $400$ femtoseconds.
The average of the net force between the two tubes is the average friction force, $F_{f}$.
When the temperature (averaged over $0.5$ ps) of either CNT
rose above $315 K$, the simulation was abandoned and a new simulation begun
using a different, and independent, equilibrated starting configuration.
$F_{f}$ was converged by averaging over
an aggregate time domain of one or two nanoseconds.
Extensive tests were performed to verify that 
the results reported here are insensitive to 
details of these simulations such as the 
frequency of  rescaling of $v_s$, the
tolerance (of $5 \%$) on temperature changes 
, the equilibration time, 
and the suddenness with which the sliding 
velocity was imposed on the shuttle and subsequently rescaled.

Figure \ref{fig1}(b) is a plot of $F_{f}$ versus $v_s$. 
The value of $v_s$ that is used here is the 
{\em average} velocity in each simulation, and not  $v_s^0$.
As expected from Fig. \ref{fig1}(a), there 
is a massive increase in $F_f$ near 
$v_s = 1100 \;\text{ms}^{-1}$ for the (10,10)-(5,5) system.
Figure \ref{fig1}(b) also shows $F_{f}$ vs $v_s$ (scaled by a factor 
of two) for a $(5,5)$ shuttle sliding within a (chiral) $(16,2)$ CNT.
In this case the shuttle and the outer tube are incommensurate 
and there are peaks in $F_{f}$ at many different velocities. 
Although the peaks in the $F_f$ are suggestive of resonances, 
and the possibility of resonant friction between CNTs has
previously been proposed\cite{servantie2}, it will be shown that this effect
can not be explained by a resonant mechanism alone. 
Due to its relative simplicity, the analysis in the remainder
of this Letter is focussed on the $(10,10)-(5,5)$ system.

When surfaces slide over one another
the relative motion of the corrugated surface potentials
directly excites particular phonons.
A phonon excited in this way can be {\em resonantly} excited if
its frequency coincides with the ``washboard frequency'', 
$\omega_{\text{wb}}=2\pi v_s /a$, where $v_s$ is the
sliding velocity, and
$a$ is a length scale that is common to both
surfaces and that plays a role in the excitation of that 
particular phonon.
$\omega_{\text{wb}}$ can take any value depending on the value of the
 $v_s$. 

To identify which phonons are involved in the observed enhancement 
of friction the {\em displacement-dependent} velocity-velocity 
correlation function 
(i.e. $A(z_i-z_j,t_1 - t_2)=\langle v_i(t_1)v_j(t_2)\rangle$, 
where $z_i,z_j$ are the displacements along the axis of the outer nanotube
of the primitive unit cells to which atoms $i$ and $j$ belong, 
and $v_i(t_1),v_j(t_2)$ are their velocities at times $t_1$ and $t_2$), 
of the (10,10) tube
was Fourier transformed in time and space. This gives the
frequency ($\omega$) and wavevector ($k = 2\pi/\lambda$) dependence
of its phonon power spectrum. Figure \ref{fig2}
shows the {\em difference} in $(\omega,k)$ space between the 
power spectra calculated at $1100 \;\text{ms}^{-1}$ and 
at $1000 \;\text{ms}^{-1}$. 
One of the two peaks in this difference spectrum is centered at 
$(\omega,\lambda)=(146 \;\text{cm}^{-1},\infty)$ 
and the other, less symmetric peak, has maximum intensity 
at $(\omega,\lambda)=(140 \;\text{cm}^{-1},3.68 \;\text{nm})$
and a substantial shoulder at $(137\;\text{cm}^{-1},3.16 \text{nm})$.
The inset to Fig.\ref{fig2}, demonstrates that, apart from these 
modes at $\omega \sim 140\;\text{cm}^{-1}$
that are strongly excited at $v_s=1100\;\text{ms}^{-1}$, the 
phonon power spectrum is rather insensitive to the sliding velocity.

The phonon band structure of the $(10,10)$ CNT at $0$ K was 
calculated by diagonalizing its dynamical matrix. These bands are
superimposed in Fig. \ref{fig3} on an image depicting the $k$-resolved power 
spectrum calculated from the MD simulations at $1000 \;\text{ms}^{-1}$. 
The phonon energetics are obviously different in the
$300$ K system with the shuttle present, however, some important 
conclusions can be drawn from this data. 
First of all, there is a wide range of velocities at which the washboard 
frequency coincides with phonon frequencies.
Even if attention is restricted to the radial breathing (RB) mode, 
``washboard resonances'' can occur at many velocities between
$1000 \;\text{ms}^{-1}$ and $2500 \;\text{ms}^{-1}$.
This means that resonance alone cannot account for the observed
dependence of $F_f$ on $v_s$.

At this point we postulate that increased friction occurs when there
is resonant excitation of a phonon in the outer nanotube 
that moves at close to the velocity of the shuttle.
In other words, the group velocity, $v_g$, of the phonon that is 
excited by the sliding motion is close to the value of $v_s$
required for a match between the phonon frequency, $\omega(k)$,  and
 $\omega_{\text{wb}}$, i.e.
the following two criteria are simultaneously met: 
(i) $v_g(k)= d\omega(k)/dk \approx v_s$ and, 
(ii)$\omega_{\text{wb}}(v) \approx \omega(k) \Rightarrow v_{\text{wb}} = 
\omega_{\text{wb}}a/2\pi \approx v_s$.
Evidence to support this hypothesis will now be provided. This evidence
consists of a demonstration that these criteria can only be 
met at velocities close to $1100 \;\text{ms}^{-1}$ by phonons that have
frequencies and wavevectors consistent with the anomalous phonon 
excitation demonstrated in Fig.\ref{fig2}, and furthermore, that these
phonon modes are those that would, most obviously, be excited by the 
washboard mechanism - the axially symmetric radial breathing (RB) and
longitudinal acoustic (LA) modes.
The LA mode and the RB mode both have perfect axial symmetry and they
mix strongly with each other 
and with other phonon modes near $(\omega,k) = (140,0.1)$
where a band anti-crossing occurs. 

In Fig.\ref{fig3}, circles mark the positions in $(\omega,k)$ space
of the peaks seen in Fig.\ref{fig2}. Four other points are marked 
with boxes. At these points criteria (i) and (ii) are both met. 
Criteria (i) and (ii) are also met at a number of other points\cite{velcriterion}, 
however, none of these other phonons either have axial symmetry or group velocities
in the range $ 950 \text{ms}^{-1} < v_g <  1250 \text{ms}^{-1}$. 
Furthermore, of these phonons, all but one (for which $\omega \approx 128 \text{cm}^{-1}$) 
have frequencies outside the range $120 \text{cm}^{-1} < \omega <  160 \text{cm}^{-1}$.
The four marked points in Fig. \ref{fig3} mostly have the character
of RB modes with some LA character in the points closest to the band anti-crossing.

The agreement between two of these four phonons 
and the simulation results presented in figures \ref{fig1} 
and \ref{fig2} is remarkably good given that they have been 
calculated from the $0$ K phonons of an isolated $(10,10)$ tube.
The values of $v_s$ and $v_{\text{wb}}$ are close to $1100 \;\text{ms}^{-1}$ and they 
are close in $(\omega,k)$ space to the positions at which anomalous 
phonon excitation occurs in the MD simulations.
Two other phonons do not, at first, appear 
consistent with figures \ref{fig1} and \ref{fig2}. The first of these
has a group velocity of $\sim 500 \;\text{ms}^{-1}$ and no peak is
observed at this velocity in Fig. \ref{fig1}. However, the wavelength 
of this phonon is $\lambda \approx 24.5$ nm which is larger than the length of the 
simulation cell, and therefore it cannot be excited in the MD simulations.
The second point, at $\lambda \approx 1.5$ nm, has a group velocity close 
to $1100 \;\text{ms}^{-1}$ but no anomalous phonon excitation at $\lambda \approx 1.5$ nm 
is observed in the MD simulations. 
A possible reason for this is provided below.

The MD simulations have been repeated
with shuttles of lengths (in nm, and excluding the lengths of
the caps) 10, 8, 5.3, 2.6 and 0 (i.e. a C$_{60}$ molecule).
No enhancement of $F_f$ was observed near
$1100$ ms$^{-1}$ or at any other velocity for the C$_{60}$ molecule.
There were peaks at $v_s = 1100 \;\text{ms}^{-1}$ for all other
shuttle lengths and the heights of the peaks increased almost
linearly with the shuttle length. 
For the $2.6$ nm shuttle, there was anomalous phonon excitation 
at the smaller wavelength $\lambda \approx 3.2 - 3.7$ nm but no anomalous phonon excitation 
at $\lambda \rightarrow \infty$. As the length of the shuttle was increased, the 
$\lambda \rightarrow \infty$ peak began to grow
at the expense of the smaller wavelength peak until, for the longest shuttle,
this smaller wavelength peak had almost disappeared.
It is reasonable that short shuttles cannot excite very long
wavelength phonons and that, as the shuttle length 
increases, phonons with a larger wavelength are excited.
Furthermore, when the washboard excitation is in-phase with a 
phonon at a given point along the outer tube it must be
in anti-phase with a point a distance $\lambda/2$ away.
Therefore,  a shuttle cannot resonantly excite a phonon
whose wavelength is substantially smaller than twice its length.
This explains the shift in intensity of the anomalous
phonon excitation from shorter to longer wavelengths as the shuttle
length increases, and the absence of anomalous excitation of 
phonons at a wavelength of $\sim 1.5$ nm.

A physical mechanism is now proposed for the results of the MD simulations:
When $v_s$ is sufficiently close to the group velocity of
a RB mode of the outer nanotube that is being resonantly excited by the
relative motion of the corrugated surface potentials of the nanotubes, 
a wave packet remains in the vicinity of the shuttle for long enough 
that it is reinforced and amplified. 
The build up of a radial distortion in the outer nanotube 
that moves along with the shuttle results. This distortion, in turn, 
changes the interaction between the two nanotubes - a feedback
that makes the response of the outer nanotube to the presence of 
the shuttle strongly non-linear. 
The details of this non-linearity are likely to be 
complicated and sensitively velocity-dependent,
but the general effect is analogous to the sharp increase 
in drag experienced by airplanes travelling at close to the speed of sound. 
At low speeds, an airplane creates a disturbance in the air and sends 
out pressure pulses that travel ahead of it to separate the air flow, thereby 
allowing air to move smoothly around it. As Mach 1 is approached, 
however, pressure waves build up to form a shock front just ahead of the airplane. 
Air in the airplane's path has little warning of its arrival and so the air's
compressibility plays a crucial role in the flow dynamics.
The dynamics become strongly nonlinear, instabilities appear in the flow, 
and lots of energy is dissipated by the shock front. The drag that results is commonly
known as ``wave drag''\cite{wavedrag} and its maximum as a function of
velocity is the ``sound barrier''.  At supersonic speeds the flow 
stabilizes and the drag is reduced.

Further evidence that a similar mechanism is responsible for the enhancement of
friction between CNTs is provided in Fig.\ref{fig4} which shows
the average radius, $r_{\text{ave}}$, of the outer CNT as a function of 
displacement, $Z$, from one end of the shuttle. The average is performed 
over the azimuthal angle and over time. At velocities of $500 \;\text{ms}^{-1}$ and 
$1000 \;\text{ms}^{-1}$, $r_{\text{ave}}$ is very similar, however,
at $1100 \;\text{ms}^{-1}$ there is a large change, suggesting that a
wave builds up around the shuttle at this velocity. The fact that the {\em difference}
between $r_{\text{ave}}$ at $1100 \;\text{ms}^{-1}$ and
at $1000 \;\text{ms}^{-1}$ is of the same order of magnitude as the variations in
$r_{\text{ave}}$ at either $500\;\text{ms}^{-1}$ or $1000 \;\text{ms}^{-1}$
implies that the inter-tube interaction is strongly affected by the distortion. 
This, in turn, implies a non-linear response of the outer CNT
to the shuttle that could explain the sharp increase in the rate of 
energy dissipation.

A vast literature exists on wave-drag and related phenomena in the 
context of aeronautics\cite{wavedrag} and much of 
the theory may be adaptable to the nanoscale.
One important difference at the nanoscale, however, is that, depending on 
the atomic level structure, this effect can occur in the same 
device at multiple velocities.
Particularly for incommensurate systems, there may be
more than one length scale that is common to the contacting surfaces and
that can cause washboard resonances.
Furthermore, multiple phonons with different group velocities may satisfy 
the criteria necessary for this effect to occur.
The $(16,2)-(5,5)$ system shown in figure \ref{fig1}(b) 
is an example in which there are many peaks in $F_f$ vs $v_s$.

In conclusion, a mechanism by which huge increases in mechanical energy dissipation 
at well defined velocities can occur at the nanoscale has been demonstrated
using atomistic simulations of coaxial carbon nanotubes sliding relative to one another.
As demonstrated in Fig. \ref{fig1}(a), this effect can have
a dramatic impact on the dynamics of a nano-mechanical system.

This work was supported by National Science Foundation 
Grant No. DMR04-39768 and by the Director, Office of Science, 
Office of Basic Energy Sciences, Division of Materials 
Sciences and Engineering Division, U.S. Department of 
Energy under Contract No. DE-AC02-05CH11231.


\begin{references}
\bibitem{technical_details} 
The Brenner potential\cite{brenner} was used to model 
the covalent bonding and a $6-12$ Lennard-Jones 
potential\cite{ryckaert} to model inter-tube interactions.
\bibitem{length_check1} All the results presented are
qualitatively independent of the length, $L$, of the periodic cell.
A small and non-monotonic quantitative dependence on $L$ is observed.
\bibitem{velcriterion}
In determining whether criteria (i) and (ii) have been met, coincidence
of velocities is taken to mean that they are 
within $100 \;\text{ms}^{-1}$ of one other.
\end{references}
\end{document}